\documentclass[aps,prd,preprint,groupedaddress,showpacs]{revtex4-1}

\usepackage{epsfig}
\usepackage{graphicx}
\usepackage{amsfonts,amssymb}
\usepackage{color}

  \usepackage{slashed}
  \usepackage{url}
\usepackage{bm}
\usepackage{verbatim}
\usepackage{float}

\def\beq{\begin{equation}}
\def\eeq{\end{equation}}
\def\bea{\begin{eqnarray}}
\def\eea{\end{eqnarray}}
\def\ba{\begin{array}}
\def\ea{\end{array}}

\begin{document}


\title{Hawking radiation and interacting fields}


\author{Marco Frasca}
\email[]{marcofrasca@mclink.it}
\affiliation{Via Erasmo Gattamelata, 3 \\ 00176 Roma (Italy)}



\date{\today}

\begin{abstract}
Hawking radiation is generally derived using a non-interacting field theory. Some time ago, Leahy and Unruh showed that, in two dimensions with a Schwarzschild geometry, a scalar field theory with a quartic interaction gets the coupling switched off near the horizon of the black hole. This would imply that interaction has no effect on Hawking radiation and free theory for particles can be used. Recently, a set of exact classical solutions for the quartic scalar field theory has been obtained. These solutions display a massive dispersion relation even if the starting theory is massless. When one considers the corresponding quantum field theory, this mass gap becomes a tower of massive excitations and, at the leading order, the theory is trivial. We apply these results to Hawking radiation for a Kerr geometry and prove that the Leahy--Unruh effect is at work. Approaching the horizon the scalar field theory has the mass gap going to zero. We devise a technique to study the interacting scalar theory very near the horizon increasing the coupling. As these solutions are represented by a Fourier series of plane waves, Hawking radiation can be immediately obtained with well--known techniques. These results open a question about the behavior of the Standard Model of particles very near the horizon of a black hole where the interactions turn out to be switched off and the electroweak symmetry could be restored.
\end{abstract}

\maketitle




\section{Introduction}

The radiation emitted by a black hole, firstly computed by Hawking \cite{Hawking:1974rv,Hawking:1974sw}, represents a paradigm to refer to for quantum gravity effects. Indeed, it provided a new view for our understanding of singularities in our universe when quantum field theory comes into play. Since then, it became textbook matter (e.g. see \cite{Mukhanov:2007zz,toms}). This effect has also been obtained as a tunneling effect at the black hole horizon \cite{Parikh:1999mf}. The result that a temperature can be attached to a black hole underpins black hole thermodynamics. Recently, a connection with quasi-normal modes of the black hole was put forward \cite{Corda:2012dw,Corda:2013nza}.

Whatever was the derivation that one can find in current literature, the underlying particle theory is always assumed to be free. No interaction behind gravitational effects is generally assumed. This with the exception of a paper by Leahy and Unruh, applied to the two-dimensional case of a Schwarzschild geometry and a $\varphi^4$ term, that showed that the effect of a horizon is to switch off the interaction \cite{Leahy:1983vb}. More recently, Collini, Moretti and Pinamonti showed that Hawking radiation survives the presence of a weak interaction ($\varphi^3$) \cite{Collini:2013moa}. In the latter, perturbation theory is applied and so, eventually, the switch--off effect could well be hidden in the involved approximations. Generally, this problem is rather complicated as the Klein--Gordon equation with a potential is not separable and one has to solve the full problem. Besides, using perturbation theory rather than to try a direct approach to the full problem can make the question of the changes to the vacuum of the interacting quantum field theory milder and easier to treat.

Recently, it has been shown how to solve exactly the Klein-Gordon equation in a Kerr--Newman geometry \cite{Vieira:2014waa}. In this way, the authors were able to derive the Hawking radiation in a rather straightforward way using the technique devised by Ruffini and Damour \cite{Damour:1976jd} and Sannan \cite{Sannan:1988eh}. Having an exact solution can make easier to derive this effect also in the presence of interacting fields. Indeed, recently we obtained a set of exact solutions to the $\varphi^4$ theory showing how a massless field can become massive \cite{Frasca:2009bc}. These results were then developed further to a full quantum field theory with a very large self--interaction term \cite{Frasca:2013tma}. The idea in this paper is to use this machinery to work out a derivation of the Hawking radiation for an interacting theory when a large self--coupling is present. This grants that we can approach the horizon of a black as near as we want and yet to observe the effect of Unruh and Leahy of the switching off of the interaction in a more general context. As a bonus, we get a solution for the Kerr black hole, easy to generalize to the Kerr-Newman case, in a full analytical way for an interacting theory. This shows the Unruh--Leahy effect at work in the limit of a self-interaction with a coupling increasingly large.

The paper is so structured. In Sec.~\ref{sec1} we introduce the exact solutions to the classical $\phi^4$ theory and their quantum interpretation. In Sec.~\ref{sec2} we introduce the Kerr metric and derive the Leahy--Unruh effect for the scalar field with a generic potential in the classical case. In Sec.~\ref{sec3} we derive the Hawking radiation for this case showing the quantum counterpart of the Leahy--Unruh effect with the disappearance of the mass gap at the horizon of the black hole. Finally, in Sec.~\ref{sec4} conclusions are yielded.

\section{Classical scalar field theory}
\label{sec1}

\subsection{Exact solution}

We consider a classical scalar field $\varphi$ satisfying the equation
\begin{equation}
\label{eq:eq1}
   \partial^2\varphi+\lambda\varphi^3=j
\end{equation}
being $\lambda>0$ the (dimensionless) strength of the self-interaction and $j$ an external source. Without source, this equation has the exact solution \cite{Frasca:2009bc}
\begin{equation}
   \varphi(x)=\mu\left(\frac{2}{\lambda}\right)^\frac{1}{4}{\rm sn}(p\cdot x+\theta,-1)
\end{equation}
where $\mu$ and $\theta$ are integration constants and ${\rm sn}$ a Jacobi elliptic function, provided the following dispersion relation holds
\begin{equation}
\label{eq:disp}
    p^2=\mu^2\sqrt{\frac{\lambda}{2}}.
\end{equation}
We sketch here a proof that grants uniqueness. Let us suppose that we are looking for traveling solutions to eq.(\ref{eq:eq1}). We take
\begin{equation}
   \varphi=a\tilde\phi(k_0t-{\bm k}\cdot{\bm x})
\end{equation}
with $a$ a scale factor, and put it into the eq.(\ref{eq:eq1}) with $j=0$. We get
\begin{equation}
   \tilde\varphi''(\xi)+\tilde\varphi^3(\xi)=0 \qquad \xi=k_0t-{\bm k}\cdot{\bm x}
\end{equation}
provided we impose the dispersion relation (\ref{eq:disp}). This differential equation defines exactly the Jacobi sn elliptic function with $k^2=-1$ \cite{nist}. One can generalize immediately this approach to the case with a generic potential $V(\varphi)$. Such a solution describes a massive field notwithstanding we started from a massless field. In presence of a source, this can be treated both classically and in quantum field theory, as shown in \cite{Frasca:2013tma}. This solution can be also given in spherical coordinates, we do this for the aims of completeness. As usual, one starts neglecting the azimuthal coordinate, the exact solution to the equation
\begin{equation}
   \partial_t\varphi^2-r^{-2}\partial_rr^2\partial_r\varphi-(r^2\sin\theta)^{-1}\partial_\theta\sin\theta\partial_\theta\varphi+\lambda\varphi^3=0
\end{equation}
takes the simple form
\begin{equation}
   \varphi(t,r,\theta)=\mu\left(\frac{2}{\lambda}\right)^\frac{1}{4}{\rm sn}(p_0t-|{\bm p}|r\cos\theta+\theta,-1)
\end{equation}
and again it must be $p_0^2-|{\bm p}|^2=\mu^2\sqrt{\frac{\lambda}{2}}$. This can be Fourier expanded into the form \cite{nist}
\begin{equation}
\label{eq:FS}
   \varphi = \sum_{n=0}^\infty B_n e^{i\frac{(2n+1)\pi}{2K(-1)}(k_0t-{\bm k}\cdot{\bm x})}+c.c.
\end{equation}
being $K(-1)=1.311028777\ldots$ the complete elliptic integral of the first kind and
\begin{equation}
      B_n=(-1)^n\frac{1}{2i}\frac{e^{-\left(n+\frac{1}{2}\right)\pi}}{1+e^{-(2n+1)\pi}}.
\end{equation}
 We have infinite plane waves with a properly ``dressed'' wave vector. Now we use the series of spherical harmonics for the plane waves \cite{jack} obtaining
\begin{equation}
    \varphi = 4\pi\sum_{n=0}^\infty B_n e^{i\frac{(2n+1)\pi}{2K(-1)}k_0t}\sum_{l=0}^\infty\sum_{m=-l}^l i^l j_l(k_nr)
   Y_{lm}(\theta,\phi)Y^\ast_{lm}(\theta_k,\phi_k)+c.c.
\end{equation}
where  $(r,\theta,\phi)$ and  $(k,\theta_k,\phi_k)$ are the spherical coordinates of the vectors $\mathbf{r}$ and $\mathbf{k}_n$, respectively, the functions $j_l$ are spherical Bessel functions and $Y_{lm}$ are spherical harmonics.

\subsection{Quantum intepretation}

Following the results in \cite{Frasca:2013tma}, the Fourier series (\ref{eq:FS}) in quantum field theory represents a tower of free particles with masses
\begin{equation}
    m_n=(2n+1)\frac{\pi}{2K(-1)}\left(\frac{\lambda}{2}\right)^\frac{1}{4}\mu
\end{equation}
and each of these states is increasingly suppressed by a factor $|B_n|^2$ that is exponentially decreasing. The given exact solution is representing oscillations around a vacuum expectation value given by $\mu(2/\lambda)^\frac{1}{4}$. Being different from zero, conformal invariance of the theory is spontaneously broken. The corresponding quantum field theory developed in \cite{Frasca:2013tma} is a perturbation theory holding in the formal limit $\lambda\rightarrow\infty$. In the opposite limit, $\lambda\rightarrow 0$, no mass gap is seen to arise but a full resummation of the perturbation series would be needed for this aim. Just a finite $\lambda$ appears to be meaningful for the theory to obtain nonlinear traveling waves.

\section{Kerr metric}
\label{sec2}

The metric generated by a black hole with angular momentum per unit mass $a=J/M$ and mass (energy) $M$ is the Kerr metric \cite{Misner:1974qy}, whose line element, in the Boyer-Lindquist coordinates \cite{Boyer:1966qh}, is given by
\begin{eqnarray}
ds^{2} & = & \frac{\Delta}{\rho^{2}}\left(dt-a\sin^{2}\theta\ d\phi\right)^{2}-\frac{\rho^2}{\Delta}dr^{2}-\rho^{2}\ d\theta^{2}\nonumber\\
& - & \frac{\sin^2\theta}{\rho^2}\left[\left(r^2+a^2\right)d\phi-a\ dt\right]^{2},
\label{eq:metrica_Kerr}
\end{eqnarray}
where
\begin{equation}
\Delta=r^{2}-2Mr+a^{2},
\label{eq:parametri_metrica_Kerr}
\end{equation}
and
\begin{equation}
\rho^{2}=r^{2}+a^{2}\cos^{2}\theta.
\end{equation}
The corresponding metric tensor is
\begin{eqnarray}
g^{\mu\nu}\!=\!\left(\!
\begin{array}{cccc}
	\frac{\left(\!r^{2}\!+\!a^{2}\!\right)^{2}\!-\!\Delta a^{2}\sin^{2}\theta}{\rho^{2}\Delta  } & 0 & 0 & \frac{a\left[\! \left(\!r^{2}\!+\!a^{2}\!\right)\!-\!\Delta \!\right]}{\rho^{2}\Delta  }\\
	0 & \!-\!\frac{\Delta }{\rho^{2}} & 0 & 0\\
	0 & 0 & \!-\!\frac{1}{\rho^{2}} & 0\\
	\frac{a\left[\! \left(\!r^{2}\!+\!a^{2}\!\right)\!-\!\Delta \!\right]}{\rho^{2}\Delta  } & 0 & 0 & \!-\!\frac{\Delta \!-\! a^{2}\sin^{2}\theta}{\rho^{2}\Delta  \sin^{2}\theta}
\end{array}
\!\right)\ .
\label{eq:metric_tensor_Kerr}
\end{eqnarray}
\begin{equation}
g \equiv \det(g_{\mu\nu})=-\rho^{4}\sin^{2}\theta.
\label{eq:det_g_Kerr}
\end{equation}
We want to prove the Leahy--Unruh effect in the most general setting. So, initially, we assume a scalar field in a generic potential. This yields the following equation
\begin{equation}
\frac{1}{\sqrt{-g}}\partial_{\mu}\left(g^{\mu\nu}\sqrt{-g}\partial_{\nu}\varphi\right)+\lambda V(\varphi)=0,
\label{eq:Klein-Gordon_Kerr}
\end{equation}
being $\lambda$ the coupling, then
\begin{eqnarray}
& & \left\{\frac{1}{\rho^2\Delta}\left[\left(r^{2}+a^{2}\right)^{2}-\Delta a^{2}\sin^{2}\theta\right]\partial^2_t-\frac{1}{\rho^2}\partial_r\left(\Delta\partial_r\right)\right.\nonumber\\
& - & \frac{1}{\rho^2\sin\theta}\partial_\theta\left(\sin\theta\partial_\theta\right)-\frac{1}{\rho^2\Delta \sin ^{2}\theta}\left(\Delta -a^{2}\sin^{2}\theta\right)\partial^2_\phi\nonumber\\
& + &\left.\frac{2  a}{\rho^2\Delta }\left[\left(r^{2}+a^{2}\right)-\Delta \right]\partial^2_{\phi t}\right\}\varphi+\lambda V(\varphi)=0.
\label{eq:mov_Kerr}
\end{eqnarray}
This reduces to the Schwarzschild metric setting $a=0$, that is
\begin{equation}
\left\{\frac{r^2}{\Delta}\partial^2_t
-\frac{1}{r^2}\partial_r\left(\Delta\partial_r\right)
-\frac{1}{r^2\sin\theta}\partial_\theta\left(\sin\theta\partial_\theta\right)
-\frac{1}{r^2\sin ^{2}\theta}\partial^2_\phi\right\}\varphi+\lambda V(\varphi)=0
\end{equation}
being now $\Delta=r^2-2Mr$. Our aim is to study eq.(\ref{eq:mov_Kerr}) near the horizon. This can be accomplished by introducing the tortoise coordinate \cite{Damour:1976jd}
\begin{equation}
    \frac{dr_*}{dr}=\frac{r^2+a^2}{\Delta}
\end{equation}
that near the horizon $r\rightarrow r_+$, being $r_\pm=M\pm(M^2-a^2)^\frac{1}{2}$, takes the form
\begin{equation}
    r_*\approx\frac{1}{2\kappa}\ln\left(\frac{r}{r_+}-1\right)
\end{equation}
with
\begin{equation}
   \kappa=\frac{1}{2}\frac{r_+-r_-}{r_+^2+a^2}
\end{equation}
the surface gravity. One gets
\begin{equation}
   \frac{1}{\rho^2}\partial_r\left(\Delta\partial_r\right)=
	 \frac{1}{\rho^2}\frac{r^2+a^2}{\Delta}\partial_{r_*}(r^2+a^2)\partial_{r_*}=
	 \frac{1}{\rho^2}\frac{(r^2+a^2)^2}{\Delta}\partial_{r_*}^2+\frac{2r}{\rho^2}\partial_{r_*}
\end{equation}
then
\begin{eqnarray}
& & \left\{\frac{1}{\rho_+^2\Delta}\left[\left(r^{2}_++a^{2}\right)^{2}-\Delta a^{2}\sin^{2}\theta\right]\partial^2_t-\frac{(r_+^2+a^2)^2}{\rho_+^2\Delta}\partial_{r_*}^2\right.\nonumber\\
& - & \frac{1}{\rho_+^2\sin\theta}\partial_\theta\left(\sin\theta\partial_\theta\right)-\frac{1}{\rho_+^2\Delta \sin ^{2}\theta}\left(\Delta -a^{2}\sin^{2}\theta\right)\partial^2_\phi\nonumber\\
& + &\left.\frac{2a}{\rho_+^2\Delta }\left[\left(r_+^{2}+a^{2}\right)-\Delta \right]\partial^2_{\phi t}\right\}\varphi+\lambda V(\varphi)=0.
\label{eq:mov2_Kerr}
\end{eqnarray}
being $\rho_+=r_+^2+a^2\cos^2\theta$. Now, near the horizon, we have $\Delta\approx(r_+-r_-)(r-r_+)$ and some terms dominate into the equation. But the non-linear term has $\lambda(\theta)=\lambda\frac{\rho_+^2}{\left(r^{2}_++a^{2}\right)^{2}}$ and so it enters like $\lambda=\lambda(\theta)\Delta$ and $\Delta\rightarrow 0$ as $r\rightarrow r_+$. So, very near the horizon, particles have only interactions with the gravitational field and the standard theory for Hawking radiation is recovered. As already said, this result should be expected in view of the analysis by Unruh and Leahy \cite{Leahy:1983vb} in two dimensions and also for accelerating frames by Unruh and Weiss \cite{Unruh:1983ac}. The interaction appears to be switched off very near the horizon of the black hole.

This result implies that a broken conformal invariance is recovered near the horizon of a black hole and the mass gap of the theory goes to zero returning to massless states. In turn, this implies that, near the horizon, the tunneling view of the Hawking radiation \cite{Parikh:1999mf} will keep holding independently if we started considering an interacting theory far from the horizon.  

To have an understanding of the theory and the fate of the mass gap, we take $\lambda$ increasingly large as $\Delta$ goes to zero. This permits to consider an interacting theory approaching the horizon. So, let us consider the limit case $\lambda\rightarrow\infty$ such that $\lambda\Delta=\tilde\lambda$ is finite as $r\rightarrow r_+$. The equation reduces to
\begin{eqnarray}
& & \left\{\frac{1}{\rho_+^2}\left(r^{2}_++a^{2}\right)^{2}\partial^2_t-\frac{(r_+^2+a^2)^2}{\rho_+^2}\partial_{r_*}^2\right.\nonumber\\
& - & \frac{1}{\rho_+^2 \sin ^{2}\theta}\partial^2_\phi\nonumber\\
& + &\left.\frac{2a}{\rho_+^2 }\left(r_+^{2}+a^{2}\right)\partial^2_{\phi t}\right\}\varphi_0+\tilde\lambda V(\varphi_0)=0.
\label{eq:mov3_Kerr}
\end{eqnarray}
We will show, in the Appendix, how good is this approximation by numerical solving this equation in a simplified case. But we just point out that this is a property of partial differential equations with a large parameter used to do perturbation theory \cite{Frasca:2007id,Frasca:2007kb}. 

In order to solve this equation, we introduce the following change of variables and specialize $V(\varphi)=\varphi^3$. One has
\begin{equation}
   \tilde\phi=(r_+^2+a^2)^\frac{1}{2}\phi \qquad g=\tilde\lambda\frac{\rho_+^2}{(r_+^2+a^2)^2}=\lambda g(r,\theta)
\end{equation}
so that
\begin{equation}
\left\{\partial^2_t-\partial_{r_*}^2-\frac{1}{\sin ^{2}\theta}\partial^2_{\tilde\phi}+\frac{2a}{(r_+^2+a^2)^\frac{1}{2}}\partial^2_{\tilde\phi t}\right\}\varphi_0+\lambda g(r,\theta)\varphi_0^3=0.
\label{eq:mov4_Kerr}
\end{equation}
with $g(r,\theta)\rightarrow 0$ for $r\rightarrow r_+$ and $\lambda\rightarrow\infty$. With the approximation shown in the Appendix, this equation can be immediately solved to give
\begin{equation}
\label{eq:exsol}
    \varphi(r_*,\tilde\phi,t)=\mu\left(\frac{2}{\lambda\ g(r,\theta)}\right)^\frac{1}{4}{\rm sn}\left(k_0t-k_*r_*-k_{\phi}(r_+^2+a^2)^\frac{1}{2}\phi+\chi,-1\right)
\end{equation}
where $\mu$ and $\chi$ are arbitrary integration constants and we used the definition of $\tilde\phi$. This holds provided the following dispersion relation holds
\begin{equation}
     \left(k_0^2-k_*^2-2ak_0k_\phi-\mu^2\sqrt{\frac{\lambda\ g(r,\theta)}{2}}\right)(1-\cos^2\theta)-k_\phi^2(r_+^2+a^2)=0.
\end{equation}
The limit cases with $\theta=0$ is obtained just by setting $k_\phi=0$ and the dispersion relation is now
\begin{equation}
\label{eq:dsr}
     k_0^2-k_*^2-\mu^2\sqrt{\frac{\lambda\ g(r,0)}{2}}=0.
\end{equation}
The mass gap very near the horizon for a strongly coupled field is then
\begin{equation}
   m_0^{(g)}(r,\theta)=m_0\left[\mu^{-2}\Delta\frac{\rho_+^2}{(r_+^2+a^2)^2}\right]^\frac{1}{4}.
\end{equation}
being
\begin{equation}
   m_0=\mu\left(\frac{\lambda}{2}\right)^\frac{1}{4}
\end{equation}
the mass gap in a Minkowski space--time. This is seen to go to zero like $(r-r_+)^\frac{1}{4}$ as discussed above. Moving through the horizon the mass becomes imaginary and the role of energy and momenta appears interchanged. This is seen by the change of sign in $\lambda$ in eq.(\ref{eq:mov4_Kerr}). It is easy to verify that, when the mass gap goes to zero, eq.(\ref{eq:exsol}) becomes a solution to the standard wave operator $\partial_t^2-\partial_{r_*}^2$ with the corresponding dispersion relation $k_0^2-k_*^2=0$ in the radial case.

\section{Hawking radiation}
\label{sec3}

Very near the horizon, the interaction is switched off and the theory becomes free. All the standard machinery to derive Hawking radiation applies (e.g. see \cite{Vieira:2014waa} or \cite{Parikh:1999mf}). Looking for purely radial solutions (here $\phi=0$ and $\theta=\pi/2$), we have
\begin{equation}
    \varphi(r_*,t)=\mu\left(\frac{2}{\lambda\ g(r,\pi/2)}\right)^\frac{1}{4}{\rm sn}\left(k_0t-k_*r_*,-1\right)
\end{equation}
provided the dispersion relation (\ref{eq:dsr}) holds (we fixed $\chi=0$). Using the Fourier expansion of the sinoidal Jacobi function, one has
\begin{equation}
\label{eq:FSr}
   \varphi(r_*,t) = \sum_{n=0}^\infty B_n^* e^{-i\frac{(2n+1)\pi}{2K(-1)}(k_0t-k_*r_*)}+c.c.=
	 \sum_{n=0}^\infty B_n e^{i\frac{(2n+1)\pi}{2K(-1)}\left[k_0t-k_*\frac{1}{2\kappa}\ln\left(\frac{r}{r_+}-1\right)\right]}+c.c.	
\end{equation}
that is
\begin{equation}
\label{eq:FSr1}
   \varphi(r_*,t) = 
	 \sum_{n=0}^\infty B_n^* e^{-i\frac{(2n+1)\pi}{2K(-1)}k_0t}\left(\frac{r}{r_+}-1\right)^{ik_*\frac{1}{2\kappa}\frac{(2n+1)\pi}{2K(-1)}}+c.c.
\end{equation}
This has the expected form \cite{Damour:1976jd}, being a weighted sum of wavefunctions of free particles in the Kerr metric. Indeed, introducing the Eddington--Finkelstein coordinate $v=t+\hat r$ and $\hat r=k_*r_*/k_0$, this becomes
\begin{equation}
\label{eq:FSr2}
   \varphi(r_*,t) = 
	 \sum_{n=0}^\infty B_n^* e^{-i\frac{(2n+1)\pi}{2K(-1)}k_0v}\left(\frac{r}{r_+}-1\right)^{ik_*\frac{1}{\kappa}\frac{(2n+1)\pi}{2K(-1)}}+c.c.
\end{equation}
that is a weighted superposition of solutions in the standard form \cite{Zhang:2005gja} with weights $B_n$. We note that now one has a tower of excitations with a mass spectrum
\begin{equation}
   m_n^{(g)}=(2n+1)\frac{\pi}{2K(-1)}m_0\left[\mu^{-2}(r_+-r_-)(r-r_+)\right]^\frac{1}{4}.
\end{equation}
Each state has a thermal radiation spectrum but this is exponentially damped at increasing $n$ as $|B_n|^2$. The mass gap is seen to go to zero as the horizon at $r_+$ is approached. The temperature is the usual $T=\kappa/2\pi$ while the Boltzmann factors are $e^{-\frac{(2n+1)\pi}{2K(-1)}\frac{k_*}{T}}$ and we see that appear heavily damped at increasing $n$. Notice that $k^*=\sqrt{k_0^2-m_0^{(g)2}}$ in agreement to eq.(\ref{eq:dsr}). The technique described in \cite{Sannan:1988eh} complete the derivation of the thermal spectrum as
\begin{equation}
   P(k_0)=\sum_{n=0}^\infty |B_n|^2\frac{1}{e^{\frac{(2n+1)\pi}{2K(-1)}\frac{k_*(k_0)}{T}}-1}.
\end{equation}

\section{Conclusions}
\label{sec4}
We have shown that the Leahy--Unruh effect, the switch-off of the interaction near the horizon of a black hole, applies generally to a scalar theory. We have considered the case of a strongly coupled field theory to move the study very near the horizon. The generation of the mass gap, that in quantum field theory produces a tower of massive excitations, does not hinder hawking radiation in view of the protection due to the Leahy--Unruh effect. On the other side, this open a more general problem to extend this analysis to other interacting theories and, more generally, to see if the Standard Model of particle theory turns out a completely massless theory near the horizon of a black hole restoring the broken electroweak symmetry. This could be an important effect connecting quantum field theory and gravity on a same footing as the Hawking radiation itself and so it is worth an in-depth study. 



\appendix*
\section{Strongly coupled PDEs}

We analyze the simpler partial differential equation
\begin{equation}
\label{eq:num}
   \partial_t^2\varphi-\partial_x^2\varphi+\lambda x\varphi^3=0
\end{equation}
that is identical to eq.(\ref{eq:mov3_Kerr}) without the angular part and after a proper redefinition of the constants. In view of the study in ref.\cite{Frasca:2007id,Frasca:2007kb}, we observe that a gradient expansion can be applied when the limit $\lambda\rightarrow\infty$ is taken and so, we are able to solve the equation exactly yielding \cite{Frasca:2009bc}
\begin{equation}
   \varphi(x,t)=\mu(x)\left(\frac{2}{\lambda x}\right)^\frac{1}{4}{\rm sn}\left(\mu(x)\left(\frac{\lambda x}{2}\right)^\frac{1}{4}t+\theta(x),-1\right).
\end{equation}
We apply this exact equation to the particular case of the boundary conditions $\varphi(0,t)=\varphi(1,t)=0$, $\dot\varphi(x,0)=0$, $\varphi(x,0)=x^2-x$. These conditions have been chosen starting from a preceding case study \cite{Frasca:2005sx}. The analytical solution takes the form
\begin{equation}
\label{eq:ana}
   \varphi(x,t)=(x^2-x)\ {\rm sn}\left((x^2-x)\left(\frac{\lambda x}{2}\right)^\frac{1}{2}t+x_0,-1\right)
\end{equation}
being $x_0={\rm arccn}(0,-1)=K(-1)$ so to have ${\rm sn}(x_0,-1)=1$. Comparison between numerical and analytical values yields the following picture
\begin{figure}[H]
\begin{center}
\includegraphics[angle=0, width=\textwidth]{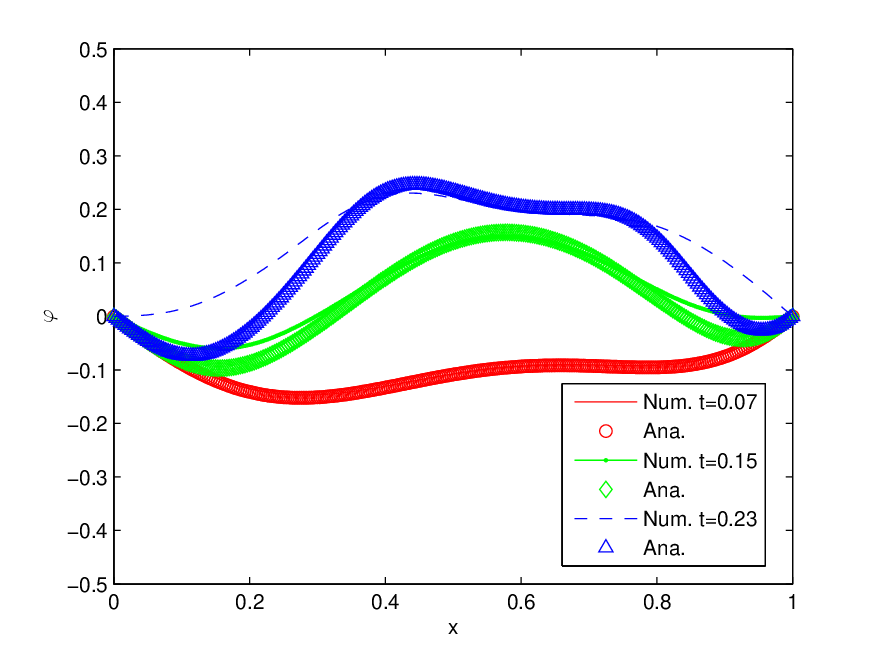}
\caption{\label{fig:fig0} Numerical solutions to eq.(\ref{eq:num}) at fixed times compared to eq.(\ref{eq:ana}) for $\lambda=10000$. Time runs from 0 to 0.3. All variables are meant dimensionless.}
\end{center}
\end{figure}
The agreement is really good and worsens, nearing the borders, as the time increases in agreement with expectations \cite{Frasca:2007id,Frasca:2007kb}.



\end{document}